\documentclass[twocolumn,aps,pre,amsmath,showpacs]{revtex4}
\usepackage{graphicx}
\usepackage{dcolumn}%
\usepackage{bm}%
\usepackage{amssymb}
\usepackage{amsmath}
\usepackage{color}

\newcommand{\br}{\mathbf{r}}

\def\be{\begin{equation}}
\def\ee{\end{equation}}

\begin{document}

\title{Non-affine deformation of inherent structure as static signature 
of cooperativity in supercooled liquids}
\author{Emanuela Del Gado, Patrick Ilg, Martin Kr\"oger and 
Hans Christian \"Ottinger}
\affiliation{ETH Z\"urich, Polymer Physics, CH-8093 Z\"urich, Switzerland}
\date{\today}
\begin{abstract}
We unveil the existence of non-affinely rearranging regions in the inherent 
structures (IS) of supercooled 
liquids by numerical simulations of two- and three-dimensional model glass 
formers subject to static shear deformations combined with local energy 
minimizations. In the liquid state IS, we find a broad distribution of 
rather large rearrangements which are correlated only over 
small distances. 
At low temperatures, the onset of the cooperative dynamics corresponds 
to much smaller displacements correlated over larger distances. 
This finding indicates the presence of non-affinely rearranging domains of 
relevant size in the IS deformation, which can be seen as the static counterpart of 
the cooperatively rearranging regions in the dynamics. 
This idea provides new insight 
into possible structural signatures of slow cooperative dynamics of 
supercooled liquids and supports the connections with elastic 
heterogeneities found in amorphous solids.     
\end{abstract}

\maketitle

When a liquid is cooled down to its glass transition temperature, particle
motion slows down enormously and becomes highly cooperative \cite{coop1}: 
The more the system becomes glassy, the more the relaxation requires 
cooperative rearrangements of a large 
number of particles. Experimental and theoretical investigations as 
well as simulation results on dynamical heterogeneities strongly support 
the presence of cooperatively rearranging regions of growing size
\cite{coop1,coop2,coop3,kob-binder}. 
However, it is still debated whether and how the onset of such 
cooperative dynamics can take place without any apparent or 
straightforward connection to structural changes. 
Such links have been searched for in various directions. 
For example, different connections between the local propensity of 
particles to motions and the underlying structural features of the system 
have been investigated, but they still remain quite elusive \cite{asaph}.

An alternative (and so far complementary) description of the dynamics of
supercooled liquids is based on the way the system explores its potential
energy surface \cite{is1}. When the system enters the 
supercooled regime, the presence of local minima, or inherent structures (IS), 
turns out to be extremely important and to produce complex trajectories in the 
potential energy landscape:
The time evolution of the system samples the basins associated to the 
local minima, or groups of such basins, on a relatively short time scale, 
whereas transitions between different groups of basins separated by 
major energy barriers take a much
longer time and imply a slow, aging process, during which the system is
out-of-equilibrium. Such a description, however, does not fully account for the 
cooperative dynamics associated to dynamical heterogeneities. Yet, possible
connections have been discussed \cite{is1} and  
recent works have gathered additional insights 
\cite{ashwin,stariolo,coslo,capa,mossa,mosaic}.

In the present paper, 
we report about a signature of the onset of cooperative dynamics in the 
structural features of supercooled liquids from a novel 
perspective. 
The approach is motivated by a recent theory 
\cite{hco_glasses} - based on a general framework 
of nonequilibrium thermodynamics - which relates the response 
of a glassy system to an applied deformation to the corresponding change 
of its IS. In contrast to previous approaches, here it 
is suggested that the {\em reversible} part of the dynamics changes 
considerably when
approaching the glass transition. Since the reversible dynamics basically
mirrors the space transformation behavior of the system, the suggestion
in \cite{hco_glasses} guides the way toward investigations of {\em static} 
rather
than dynamical properties. Accordingly, the most
important feature of IS transformation is a hindrance to
affine deformation: The typical linear size of the regions where non-affine 
deformations take place is roughly the particle size in the liquid and 
is expected to become larger upon approaching the glass transition.

We explore these theoretical considerations by computer 
simulations of different model systems for supercooled liquids subject 
to static shear deformations with small amplitudes. 
In the liquid phase,  
a broad distribution of rearrangements is found, 
centered around a rather large mean value. 
When the system enters the supercooled regime, instead, 
its IS displays the onset of an enhanced collective behavior with 
much smaller displacements.
This feature, which does not depend on the particular model or 
deformation considered, has been detected here for the first time and
indicates the existence of non-affinely rearranging 
regions of relevant size in the IS of supercooled liquids. 
The presence of such collective rearrangements
in the IS response is strongly evocative of the cooperativity 
characterizing the dynamics, and it is in fact observed in the same 
range of temperatures. 
On this basis, we propose that these non-affinely rearranging regions 
are the IS counterpart of the cooperatively rearranging regions 
observed so far only in the dynamics.
 
{\em Methods and numerical simulations.} We consider three 
well established model systems for supercooled liquids: 
({\em a}) The three-dimensional (3D) 
binary Lennard-Jones mixture of Ref.~\cite{kob-andersen}; 
({\em b}) The 3D soft spheres binary mixture studied in Ref.~\cite{caruzzo}; 
({\em c}) The 2D soft spheres binary mixture in 
Ref.~\cite{harrowell99}. 
In all three models, the densities have been chosen such that 
crystallization is prevented. 
Upon cooling, these systems therefore go from the liquid to the supercooled 
and glassy regime. 
In order to investigate the response of IS to an applied deformation 
we have designed the following procedure \cite{hco_glasses}.
First, starting from a particle configuration $X=\{\br_{i}\}$ 
equilibrated at a given temperature $T$, 
we deform it affinely $\br_i\to\br_i^{\rm d}$ 
and subsequently find the inherent structure of the deformed 
configuration $X^{\rm dq}=\{\br_i^{\rm dq}\}$. For the energy 
minimization we use a conjugate gradient algorithm.
Second, starting from the same initial configuration $X$, we first 
find the IS corresponding to the initial configuration 
$X^{\rm q}=\{\br_i^{\rm q}\}$ and subsequently 
apply the same affine deformation to obtain  
$X^{\rm qd}=\{\br_i^{\rm qd}\}$. 
All subsequent analysis is based on the comparison between the two 
configurations $X^{\rm dq}$ and $X^{\rm qd}$, which is quantified by the 
mismatch, i.e.~the non-affine displacement field 
$\{{\bf d}_i=\br_i^{\rm dq}-\br_i^{\rm qd}\}$.
In the limit of a flat energy landscape, 
$\{{\bf d}_i\}$ is identically zero.
By lowering the temperature into the supercooled regime, we expect 
the presence of basins and barriers in the energy landscape to 
introduce collective contributions to the displacement field. 
We carefully chose the deformation amplitudes sufficiently large to make 
the system typically leave a basin. 
Supplementary information is given in the EPAPS document \cite{epaps}.
We have considered two different kinds of deformation: ({\em i}) The 
spatially modulated shear deformation $\br_i \to \br_i^{\rm d} = 
\br_i + \gamma \sin(k_n y_i) {\bf e}_x$, where $\gamma$ denotes the 
maximum amplitude of the deformation and $k_n=n\pi/L$ the wave-vector with 
$L$ the size of the simulation box; ({\em ii}) The homogeneous shear 
$\br_i \to \br_i^{\rm d} = \br_i + \gamma y_i {\bf e}_x$, 
where Lees-Edwards boundary conditions are used. We have applied the 
deformations ({\em i}) and ({\em ii}) to the systems 
({\em a}), ({\em b}) and ({\em c}) with $\gamma$ varying between 
$10^{-4}$ and $10^0$.
All quantities are given in reduced Lennard-Jones units. 
For all the systems we have prepared $20$ to $50$ independent samples of $N$ 
particles ($N$ varying from $500$ to $8000$) \cite{epaps} 
which have been carefully equilibrated \cite{long}. All quantities 
discussed are averaged over the independently prepared 
samples and the error bars obtained from sample-to-sample fluctuations.
Some simulations were performed using the LAMMPS code \cite{lammps}.
The following analysis concerns temperatures $T$ in the range from the 
high $T$ Arrhenius dependence of dynamic quantities to the onset of caging and 
cooperative dynamics.

{\em Results.}
\begin{figure}
\begin{minipage}{0.49\linewidth}
\includegraphics[width=1.\linewidth,height=1.\linewidth]{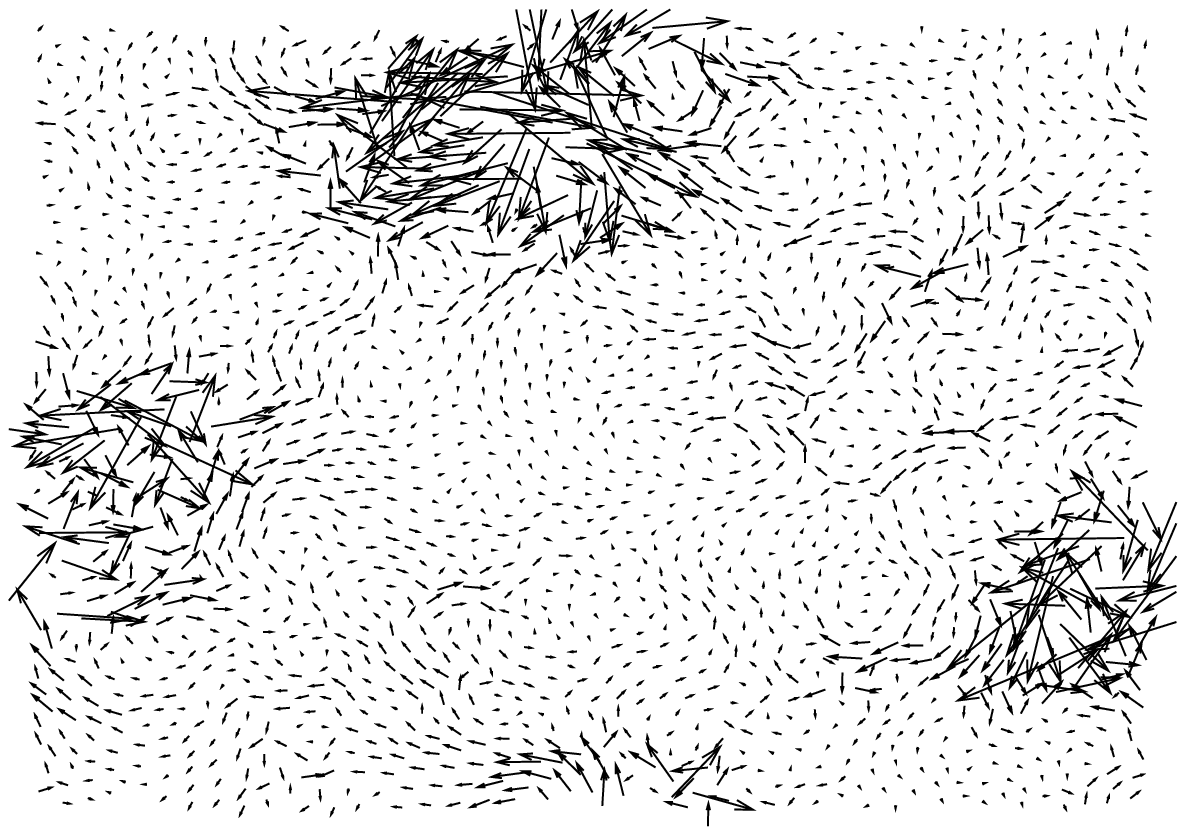}
\end{minipage} 
\begin{minipage}{0.49\linewidth}
\includegraphics[width=1.\linewidth,height=1.\linewidth]{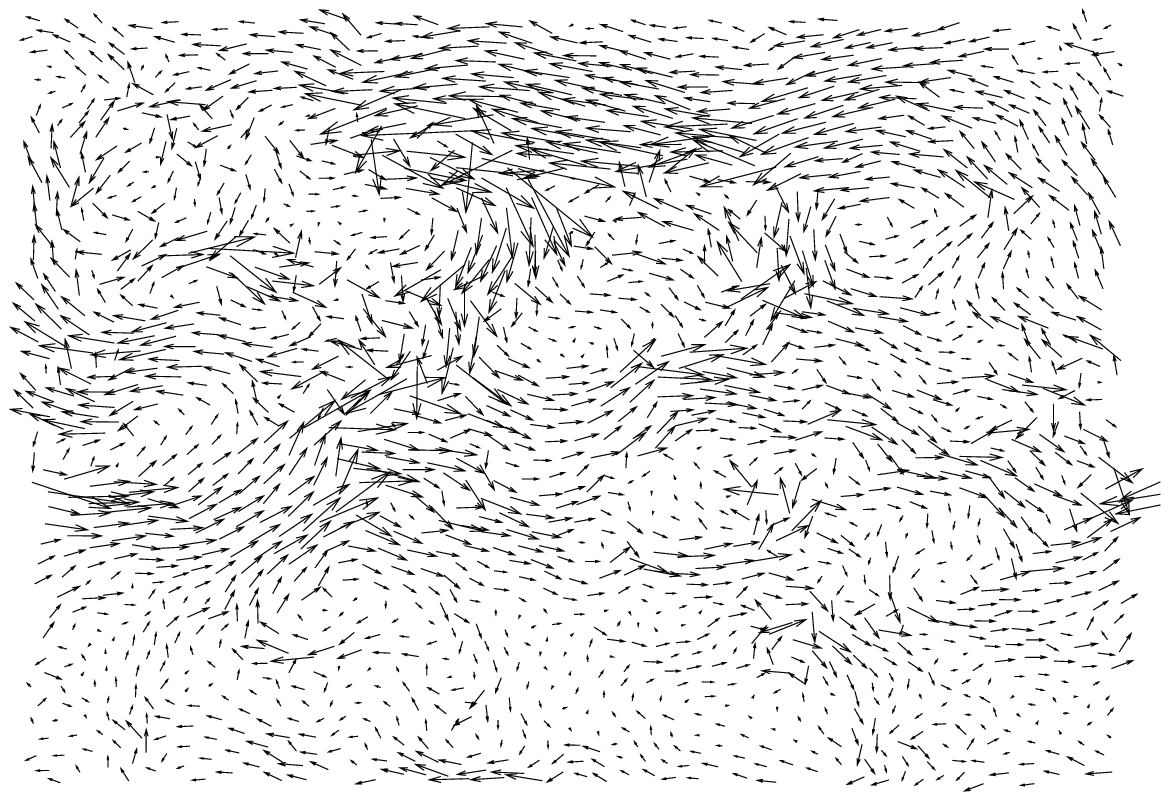}
\end{minipage}
\caption{Mismatch field $\{{\bf d}_i\}$ for
the 2D soft spheres binary mixture of Ref.~\cite{harrowell99}
at $T=1.0$ (left) and $T=0.46$ (right) for homogeneous shear deformation
with amplitude $\gamma=0.001$.
For better visibility, the lengths of the vectors are multiplied 
by a factor 14 and 600
for $T=1.0$ and $T=0.46$, respectively.}
\label{fig1}
\end{figure} 
A first, qualitative understanding can be obtained via a simple visualization 
of the mismatch field $\{{\bf d}_i\}$ at different temperatures 
and for different amplitudes of deformation. 
In Fig.~\ref{fig1}, $\{ {\bf d}_i \}$ vectors are plotted 
for the system ({\em c}) and 
deformation ({\em ii}) 
at high temperature $T=1.0$ (left) and low temperature $T=0.46$ (right). 
The length of the vectors are suitably enlarged for 
clarity.
At high $T$, for a fixed amplitude of deformation, 
the mismatch vectors display a relatively large magnitude 
as well as uncorrelated directions. This indicates that, at these 
temperatures, the local changes to the IS induced by our procedure are in 
fact dominated by the thermal fluctuations of the system. At low 
temperatures, instead, $\{{\bf d}_i\}$ have much smaller magnitudes 
for the same $\gamma$ (see figure caption for details) but, at the same 
time, they appear rather spatially correlated in magnitude and in direction. 
These two 
kinds of behavior are ubiquitous in our numerical study and appear 
systematically for a wide range of $\gamma$ values \cite{long}. 
Interestingly enough, in all cases considered, 
the transition from the one to the other behavior 
is coupled to the onset 
of the supercooled regime (for example, for the system ({\em c}) 
in Ref.~\cite{harrowell99} the onset of the glassy dynamics 
can be set around $T\approx 0.46$). 
These findings support the idea that a qualitative change of the 
response of IS to deformation is strongly correlated to the onset 
of the cooperative dynamics \cite{ashwin}. 

A first, simple characterization is given in terms of the mean length of 
the mismatch 
vectors $l_{d}=\langle N^{-1}\sum_{i}{\bf d}_{i}^{2}\rangle^{1/2}$. 
In Fig.~\ref{fig2}, $l_{d}(\gamma,T)/l_{d}(\gamma,T\!=\!1)$, 
obtained in the system ({\em b}) using the deformation ({\em i}) 
of amplitude $\gamma$ and wave vector $k_1=\pi/L$, is plotted as a function 
of the temperature $T$.  
For sufficiently small values of $\gamma$, 
the relative global magnitude of the mismatch field shows a 
rather steep decrease with decreasing temperature. 
\begin{figure}
\vspace*{0.8cm}
\includegraphics[width=1.0\linewidth]{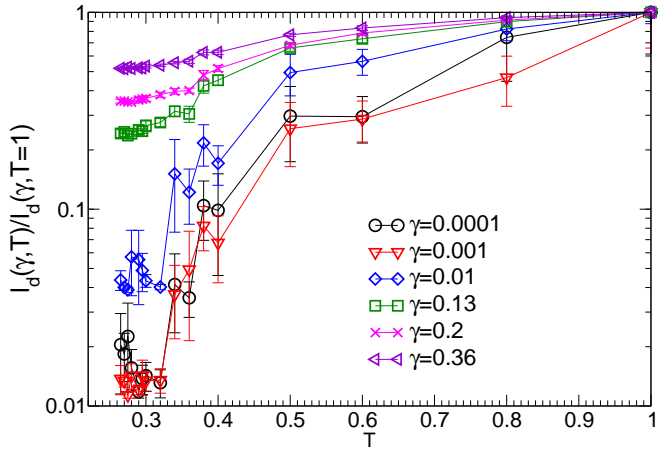}
\caption{(color online).  
Normalized mean length of mismatch field  
$l_{d}(\gamma,T)/l_{d}(\gamma,T\!=\!1)$ as a function of $T$ in the 
3D binary mixture of Ref.~\cite{caruzzo}, 
obtained with the spatially modulated deformation ({\em i})
of amplitude $\gamma$ and wave vector $k_1$.
}
\label{fig2}
\end{figure}
In fact in Ref.~\cite{ashwin}, it has also been observed
that the critical deformation amplitude to cause IS transitions strongly 
depends on temperature.
Remarkably, 
there are clearly two well distinguished temperature 
regimes for the typical length of the mismatch vectors. In addition, 
the transition from the high temperature to the low 
temperature one becomes steeper for small $\gamma$ and takes place 
at the onset of the cooperative dynamic regime ($T\simeq0.4$),
i.e.~very close to $T_{\rm eIS}$, where the IS energy starts to strongly 
depend on temperature, and well above structural arrest 
($T_{\rm MCT} \simeq 0.303$) \cite{caruzzo,epaps}.

An estimate of the fraction of particles involved in
the rearrangements described by the mismatch field 
can be obtained by the dimensionless width of the length distribution
$\Pi  = \langle N^{-1}[\Sigma_{i}{\bf d}_{i}^{2}]^2/\Sigma_{i}
[{\bf d}_{i}^{2}]^2\rangle$ (the participation ratio of 
Ref.~\cite{fabien_nonaffine1}).
In the main frame of Fig.~\ref{fig3}, $\Pi$ is plotted as a function of 
temperature $T$ for $\gamma=10^{-4},10^{-3}$ and $10^{-2}$ 
for the same system and the same wave vector as in Fig.~\ref{fig2}. 
The width $\Pi$ increases with decreasing temperature below $T=0.4$. 
In particular, for the very small values of $\gamma$ shown in the figure, 
$\Pi$ displays a rather sharp increase and 
shows that the fraction of particles involved in such non-affine 
deformation of the IS increases from $10\%$ to $40\%-60\%$. 
In order to further characterize the nature of the non-affine rearrangements 
 we have measured the average degree of correlation 
$C(r)$ between the directions of the mismatch vectors of different 
particles $i$ and $j$ separated by a distance $r$, $C(r)= 
\langle {\bf d}_{i} \cdot {\bf d}_{j} \rangle /\langle 
{\bf d}_{i}^{2} \rangle$. In the inset of Fig.~\ref{fig3}, $C(r)$, 
obtained for $\gamma=0.01$ in the system ({\em b}) with the deformation 
({\em i}), is plotted as a function of $r$. The data show that
the low temperature regime is characterized by a definitely larger degree of 
correlation at the same distance $r$. Moreover, the distribution of values of 
${\bf d}_{i} \cdot {\bf d}_{j} /\langle {\bf d}_{i}^{2} \rangle$ for pairs 
of particles separated by distance $r \pm dr$, plotted in Fig.~\ref{fig4}, 
widens strongly at low temperature, indicating that the extended 
correlation in the response of IS to deformation corresponds to a high degree 
of heterogeneity. This feature is strikingly similar to what is observed in 
the dynamics with the average time correlation and the fluctuations around 
this average, generally used to quantify dynamical heterogeneities. 
But it is observed here, in the same range of temperatures \cite{epaps}, 
in a {\em static} quantity.   
\begin{figure}
\vspace*{0.8cm}
\includegraphics[width=1.0\linewidth]{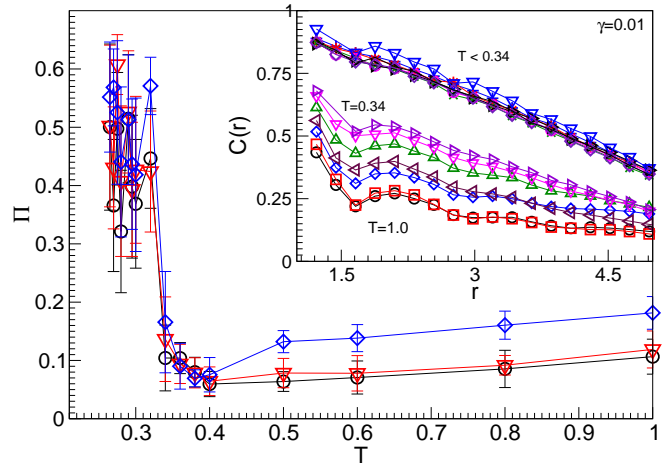}
\caption{(color online). Main frame:
The participation ratio $\Pi$ as
a function of the temperature for the 3D 
binary mixture of Ref.~\cite{caruzzo}, in the case of the spatially modulated
deformation ({\em i}) of amplitude $\gamma=10^{-4},10^{-3},10^{-2}$.
Inset: Average degree of correlation $C(r)$ of the direction
of the mismatch vectors separated by a distance $r$ for $\gamma=0.01$.
}
\label{fig3}
\end{figure}
\begin{figure}
\vspace*{0.8cm}
\includegraphics[width=1.0\linewidth]{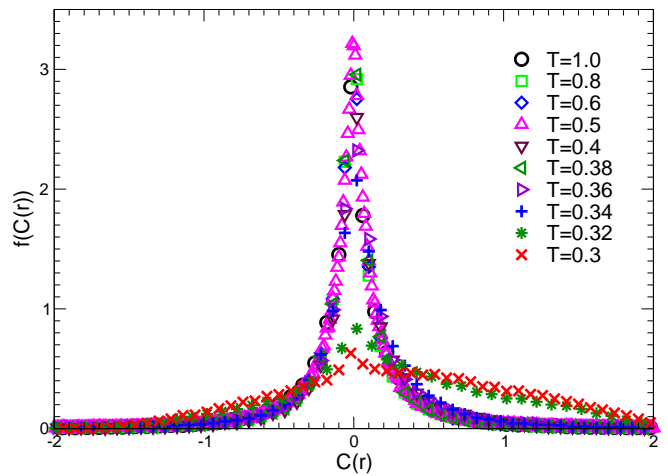}
\caption{(color online). Distributions of degree of correlation $C(r)$ 
between particles at a 
distance $r=5$ in particle diameter units. 
The same system and the same spatially modulated deformation ({\em i}) 
as in Fig.~\ref{fig3} was chosen with deformation amplitude $\gamma=0.01$.   
}
\label{fig4}
\end{figure}
On the whole, the emerging new scenario is that  
the onset of cooperativity in the real-time dynamics
corresponds to  
growth of correlated and heterogeneous domains in the non-affine 
deformations of the IS. 

The same qualitative behavior is obtained in the entire range 
$10^{-4} \leq \gamma \lesssim 0.1$ and in the systems ({\em a}) and 
({\em c}) for both deformations ({\em i}) and ({\em ii}). 
We have also found in all the systems considered that, when $\gamma\geq1.0$,  
the mismatch field always closely resembles 
the one obtained at high temperature. This indicates that  
beyond $\gamma\simeq1.0$ the amplitude of the deformation is sufficient to 
erase any distinctive feature of the low temperature structure of the IS, 
as in a sort of Lindemann criterion \cite{is1}. 
Finally, the details of temperature dependence
of the different quantities, e.g.~the crossover width or
the values of $\Pi$, depend to some extent on the 
specific system considered (e.g.~the crossover is always 
smoother in the 2D system) and on the type of deformation \cite{long}.

{\em Conclusions.}
The presented results reveal for the first time that the onset 
of the glassy dynamics corresponds to the onset of cooperative heterogeneous 
rearrangements in the response of IS to external, static deformations.
In fact we have shown that in the liquid state, as a response to the external 
deformation, the IS displays large and uncorrelated rearrangements, 
dominated by thermal fluctuations. 
Near the onset of the cooperative dynamics, the response of the IS to small 
external deformation is instead characterized by small, collective 
non-affine rearrangements involving a relevant fraction of particles 
\cite{hco_glasses}. 
To our knowledge, this feature of the IS had not been recognized until now.
It has been possible here, thanks to the procedure designed in
Ref.~\cite{hco_glasses} which strongly relies on the many-body nature of the
IS construction.
An interesting advantage of our approach is in avoiding calculations 
based on the dynamics,
which usually require very long simulation times. The extent of such an
advantage can be better quantified by systematically comparing our procedure
to the dynamics \cite{long}.

The heterogeneous domains of the IS, determined by collective non-affine 
displacements, are strongly evocative of the cooperatively rearranging regions 
observed in the dynamics and are in fact detected in the 
same range of temperatures. 
We propose therefore that they are actually the 
IS counterpart of the cooperatively 
rearranging regions. If this is the case, they would deliver a 
{\em static} correlation length which grows 
significantly in the supercooled regime, 
where the dynamic correlation length related to dynamic cooperativity 
starts to grow. 
From this point of view, this study suggests a new path for the 
investigations of structural signatures of glassy dynamics.  
A more sophisticated analysis of the non-affine displacement field, 
for example, adds useful insights in this direction 
\cite{epaps}.
Finally, our results have also interesting connections with the studies 
of elastic heterogeneities in amorphous solids 
\cite{suter,fabien_nonaffine1,fabien_nonaffine2,picard,malandro,langer04,lemaitre,papa}. 
Their typical length scale has actually been related to the non-affine part 
of the displacement field generated in response to an elastic deformation.
In Ref.~\cite{fabien_nonaffine2}, 
where a quantitative estimate of this length scale is obtained in a
realistic model of amorphous silica, the connection between 
the typical length scale of elastic 
heterogeneities in the amorphous solid and the length scale 
typical of dynamical heterogeneities (which is actually of 
comparable magnitude) in the supercooled liquid is suggested. 
Our results support this idea and offer a way to extend those studies 
to finite temperatures. 
On approaching the supercooled regime from the high temperature fluid, 
we find a signature of the onset of dynamic cooperativity in a 
purely static quantity: the non-affine, collective rearrangements of the IS  
show a striking similarity to cooperative, dynamical phenomena.

{\em Acknowledgements.} This project has been supported through EU-NSF contract
NMP3-CT-2005-016375 of the European Community.

%\documentclass[twocolumn,aps,pre,amsmath]{revtex4}
%
%\usepackage{graphicx}
%\usepackage{dcolumn}%
%\usepackage{bm}%
%\usepackage{amssymb}
%\usepackage{amsmath}
%\usepackage{color}
%\usepackage{colordvi}
%\usepackage{times}
%\usepackage{datetime}
%\usepackage{fmtcount}
%\newcommand{\kb}{k_{_{\rm B}}}
%\newcommand{\br}{\mathbf{r}}
%\newcommand{\ave}[1]{\langle #1 \rangle}

%\newcommand{\bea}{\begin{eqnarray}}
%\newcommand{\eea}{\end{eqnarray}}
%\newcommand{\mbf}[1]{\mbox{\boldmath{$#1$}}}
%\newcommand{\avee}[1]{\left\langle #1\right\rangle}
%\newcommand{\X}{\Delta}
%\newcommand{\href}{h_{\rm ref}}

%\def\be{\begin{equation}}
%\def\ee{\end{equation}}
%\def\k{{\bm{k}}}
%\def\r{{\bm{r}}}
%\def\lan{\langle}
%\def\ran{\rangle}

%\begin{document}

\setcounter{figure}{4}

\title{{Auxiliary Material}\\
Non-affine deformations of inherent structure as \\static signature of
cooperativity in supercooled liquids \\}
\author{Emanuela Del Gado, Patrick Ilg, Martin Kr\"oger and
Hans Christian \"Ottinger}
\affiliation{ETH Z\"urich, Polymer Physics, CH-8093 Z\"urich, Switzerland}
\date{\today}

\maketitle

\begin{appendix}

\section{Details of Numerical Simulations}
For the two 3D systems ({\em a}) and ({\em b}), we have performed NVE
and NVT molecular dynamics simulations of systems containing
$N=2048$ particles. For system ({\em a}) and ({\em b}), the chosen
number densities and corresponding size of the simulation box are
$\rho=1.2$, $L=11.95$ \cite{Akob-andersen} and $\rho=0.6$, $L=15.06$
\cite{Acaruzzo}, respectively. For the 2D system ({\em c}), we have
performed constrained molecular dynamics simulations in the
$NPT$-ensemble (with $N=500$, $2048$, and $8000$), with
Nos{\'e}-Hoover thermo- and barostats ensuring on average a fixed
temperature ($0.3\leq T \leq 2$) and pressure ($P=13.5$) value
\cite{Aharrowell99}. 

\section{The mismatch distribution} 

The motivation for studying the mismatch field is the
idea, presented within the thermodynamic modeling approach of
Ref.~\cite{Ahco_glasses}, that the behavior of the system under deformation
can be described using the underlying inherent structure. 
Since we here study the liquid regime, we have chosen the mismatch 
field $\{\mathbf{d}=X^{\rm dq}-X^{\rm qd}\}$ in order to subtract 
the affine part of the deformation, such that
$\mathbf{d}$ only accounts for non-affine contributions to the
deformation. For a crystalline solid, instead, the mismatch has to
be more conveniently considered with respect to $X^{\rm q}$, i.e. 
the IS of the original configuration. In this case, $\mathbf{d}$ is 
therefore a measure of the difference between two configurations of the solid. 

There are number of related and special cases which help to clarify
the meaning of the mismatch vectors $\{{\bf d}_j\}$, or equally, the
mismatch vector field ${\bf d}({\bf r})$: (A) For the special case of a 
perfectly flat energy landscape, 
$\mathbf{d}$ vanishes identically; (B) 
In a real fluid, the mismatch vector field $\mathbf{d}$ does 
not vanish even at high temperatures, in fact its magnitude 
tends to become large in that case, but it will
be orientationally correlated only over short distances and
vanish when averaged over volume elements
of a few particle diameters; 
(C) In the case of 
extremely small deformations that do not allow to overcome
even the slightest energy barriers, $\mathbf{d}({\bf r})$ is
identical to minus the imposed deformation at location ${\bf r}$.
This limit has not been reached for the range of deformation
amplitudes $\gamma$ that we have considered (see also next section);
(D) Applying rather large deformations
$\gamma\gtrsim 10^{-1}$ within our procedure (but out of focus of
the present study) gives information about sizes of inherent
structure basins, similar to the work of \cite{Aashwin}; (E) For very
low temperatures, where the initial configuration is an amorphous
solid, $X\to X^{\rm q}$, the mismatch field coincides with the
quantity studied in Ref.~\cite{Aleonforte}, where it was used in
order to extract the non-affine displacements in amorphous solids.

\begin{figure}
\includegraphics[width=8cm]{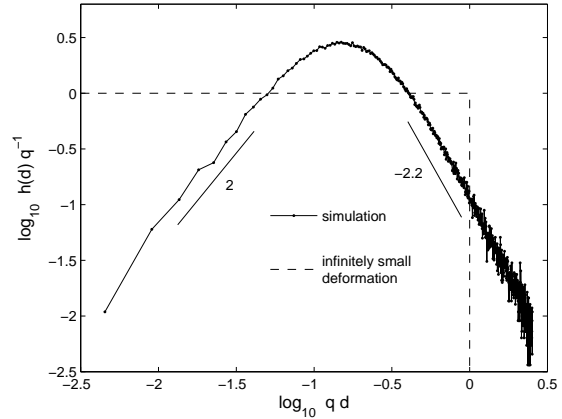}
\caption{Scaled probability distribution function of mismatch strengths,
with $q\equiv 2/(\gamma L)$, for $T=0.46$ and $\gamma=10^{-3}$ for system (c).
Also shown is the reference distribution one would observe for
extremely small deformations, as described in the text part.}
\label{addon01}
\end{figure}

\begin{figure}
\includegraphics[width=8cm]{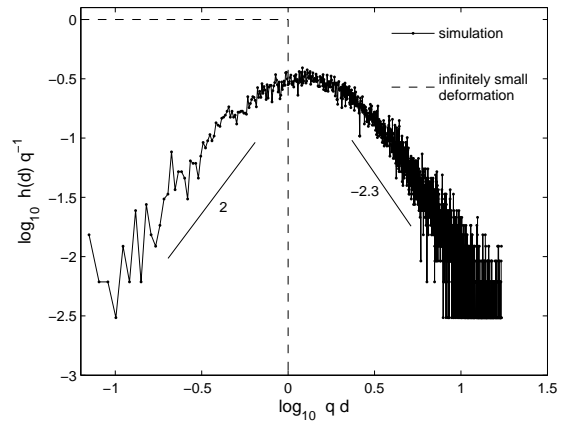}
\caption{Same as Fig.~\ref{addon01} for $T=0.8$ and
$\gamma=10^{-3}$.}
\label{addon02}
\end{figure}

From the measured mismatch fields $\{{\bf d}_j\}$ we have further
compiled ensemble averaged probability distributions of mismatch
strength; denoted as $h(d)$, where $d$ is the norm of the mismatch
vector.
For case (C) considered above,
the distribution is flat and vanishes for $d\ge \gamma L/2$
(for the case of homogeneous shear strain applied of size $\gamma$
to a homogeneous system of linear size $L$).
In Figs.~\ref{addon01} and \ref{addon02},
we quantitatively demonstrate for two representative 
temperatures and strain amplitudes $\gamma$,
that even for deformations as small as
those required for the approach presented in the manuscript, this
regime is not observed.
Moreover, the figures show that the distributions are characterized by
an `initial' (at small $\gamma$) parabolic increase with $d$, a `later' 
decrease
well described by a power law, and a finite range of displacement strengths which ensures
integrability and calculation of the moments $\Pi$ and $l_d$ presented in Figs.~2 and 3 of
the manuscript.
The occurrence of a significant fraction of values of $qd$ between $1$ and 
$10$ in Fig.~6 underlines the enormous freedom 
of rearranging individual particles in the liquid state.

The strong discrepancy between measured distributions and the
reference distribution, as shown in Figs.~\ref{addon01} and
\ref{addon02} for two different temperatures, sheds some light on
the fact that a picture of energy valleys being large enough to
accommodate small deformations without affecting the inherent
structure begins to fail at deformations small compared to the ones
needed to recognize the occurrence of the glass transition.

\section{Propensity of particles to motion}
For the model of Ref.~\cite{Acaruzzo}, to which most of the data shown in the 
paper refer, no study of dynamical heterogeneity was available in the 
literature. We have calculated the local propensity to motion $p$ as defined 
in Ref.~\cite{Aasaph} from the distribution of particle displacements after 
a time interval $\simeq 1.5 \tau_{e}$, where $\tau_{e}$ is the time at which the
intermediate scattering funcion, measured at the wave vector of the first 
peak in the structure factor, has decayed to $1/e$. The distribution is 
calculated from $N_{r}=20$ separate simulation runs over a fixed time interval,
all starting from the same particle configuration but with momenta randomly 
assigned from the Maxwell-Boltzmann distribution at a certain temperature.
In Ref.~\cite{Aasaph} it is shown that the probability distribution $f(p)$ 
becomes asymmetric and much wider at the onset of the 
supercooled regime. In Fig.~7 we have plotted the distribution of
propensity values at various temperatures: the data show the change in 
the shape of the distribution at temperatures below $T=0.4$ and higher than
$T_{\rm MCT}$. The distribution of degree of spatial correlation plotted in 
Fig.~4 of the manuscript displays a strikingly similar qualitative change in the
same range of temperature. 
\begin{figure}
\vspace{2.0cm}
\includegraphics[width=1.0\linewidth]{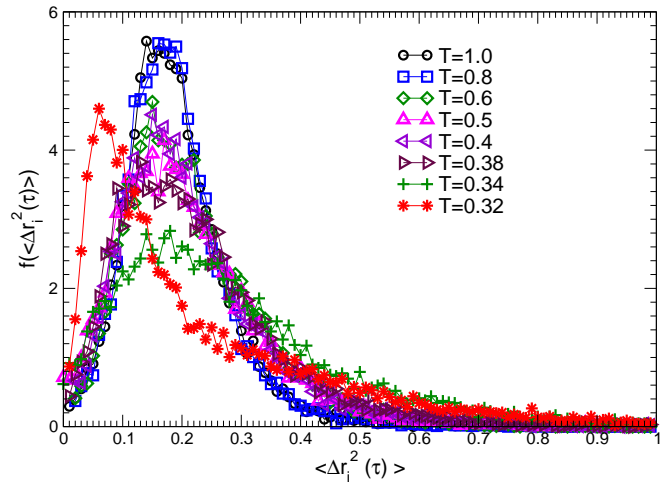}
\label{figure3}
\caption{Distribution of particle displacements (propensity to motion) 
at time $\tau \simeq 1.5 \tau_{e}$ at different temperatures.}
\end{figure}
\section{Search for correlations between distinct approaches}
One of the possible developments of this work is in the direction 
of quantitatively characterizing correlations between the cooperative domains
observed in the response of IS and the heterogeneities observed in particle 
dynamics. In principle, this would require a systematic analysis of the 
different measures of dynamical heterogeneities proposed in the literature 
\cite{Adh}, but, to get started, let us consider 
the local propensity to motion $p$ discussed in the previous secion.
A significant change of shape is also observed for the
distribution $h(d)$ of magnitudes of the mismatch vectors, as
discussed in the previous section, and even more strikingly for
the distribution of $\langle {\bf d}_i\cdot{\bf d}_j\rangle$, $f(C)$, 
see Fig.~4 of the manuscript. 
Starting from these first indications, 
we have calculated conditional probability distributions for the 
propensity and mismatch lengths. We have also evaluated, 
at different temperatures,
different correlation coefficients between propensity and
mismatch length distributions, the latter calculated for
different values of deformation amplitudes.
\begin{figure*}[ht]
\includegraphics[width=0.48\linewidth]{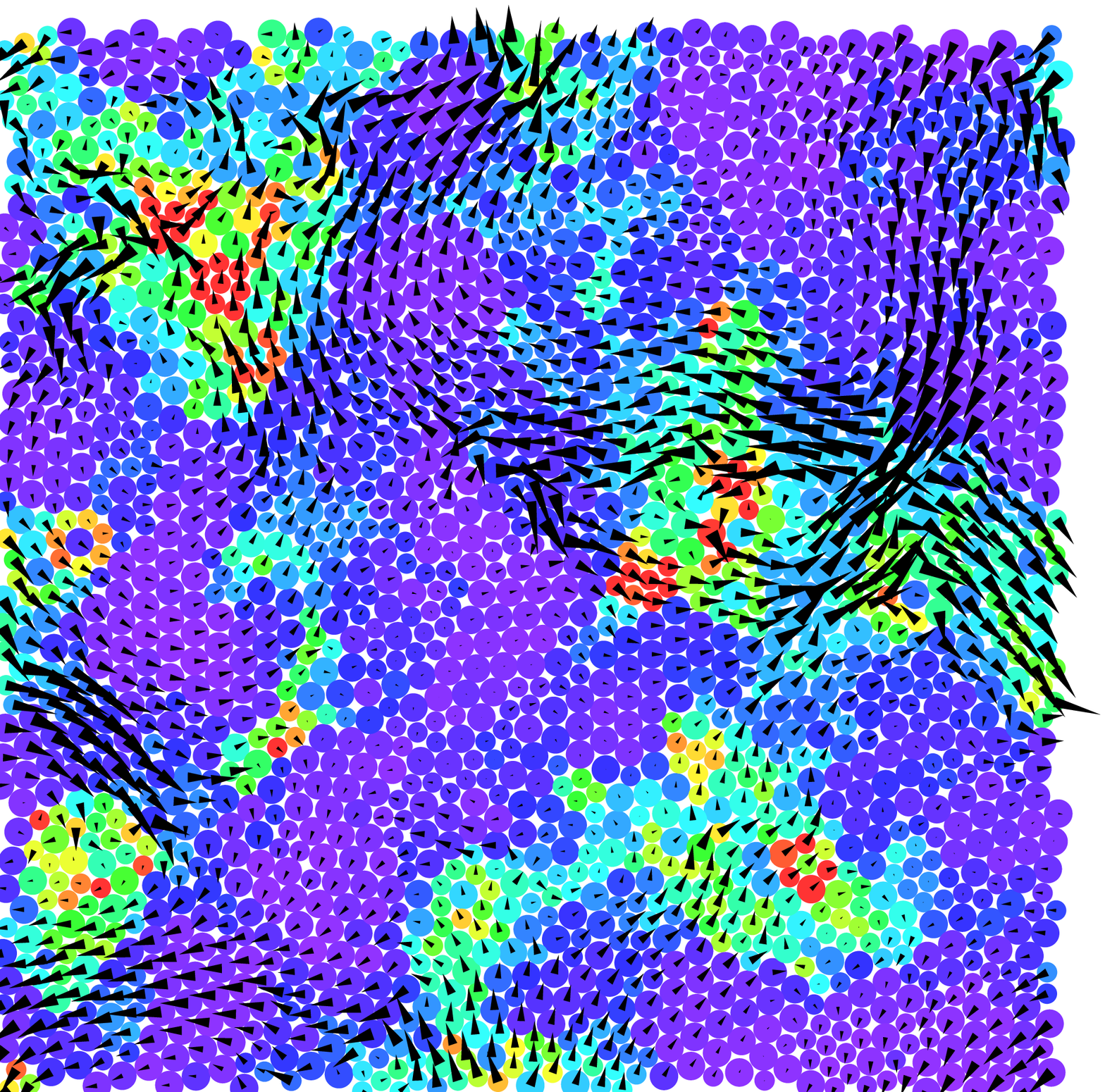}
\includegraphics[width=0.495\linewidth]{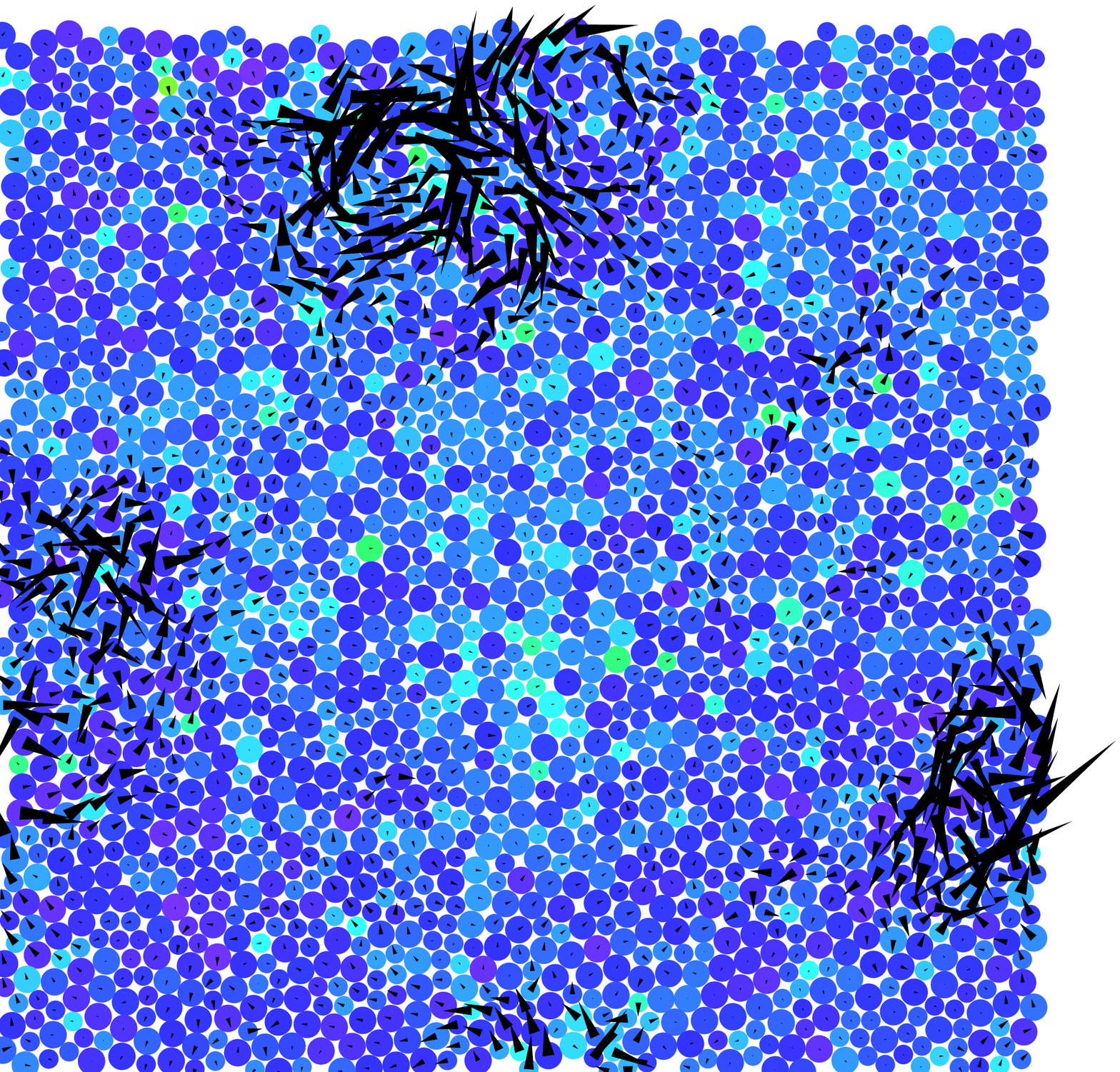}
\caption{Visualization of particle propensity and
mismatch vectors for the system ({\em c}). Increasing propensity is
color coded from violet (dark) to red (bright). Same conditions are
chosen as in Fig.~1 of the manuscript with $T=0.4$ (left) and $T=1.0$
(right).} \label{addons2}
\end{figure*}
We can rule out a clear signal of a simple and 
straightforward correlation between
$|{\bf d}_j|$ and $p_j$. This is illustrated e.g.~in Fig.~\ref{addons2},
where the propensities and mismatch vectors are visualized on top of the
particle configuration for the 2D system ({\em c}).
Correlations between particle propensity and the spatial correlation of the 
mismatch vectors give instead more encouraging indications, as also suggested 
by the cooperative nature of the mismatch vectors at low temperature.
By defining the coarse-grained mismatch field
$D(r)= \langle \mathbf{d}^{2}(r) \rangle _{j}$, where $\langle...\rangle_{j}$
indicates the average over the $N_{j}$ particles contained in the volume
$V_{j}$ of linear size $r$ \cite{Aleonforte}, and 
the analogue coarse-grained propensity field $P(r)$,
we have calculated their spatial correlation $\langle D(r)P(r)\rangle$, 
plotted in Fig.~8 for the model system ({\em b}).
Such correlations  
apparently become longer ranged at low temperatures.
The same behavior is observed in different model systems,
and for different shear amplitudes.
\begin{figure}
\vspace{0.3cm}
\includegraphics[width=1.0\linewidth]{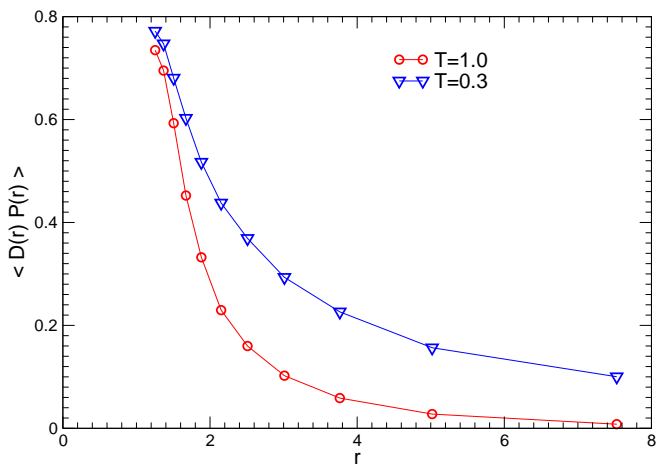}
\label{figure8}
\caption{Spatial correlations between the coarse grained fields
$D(r)$ and $P(r)$ as a function of the coarse graining distance $r$.}
\end{figure}

\end{appendix}

\end{document}